\documentclass[preprint,showpacs]{revtex4}
\usepackage{epsf}
\usepackage{latexsym}

\newcommand{\bS}{\mbox{\boldmath{$S$}}}

\newcommand{\bJ}{\mbox{\boldmath{$J$}}}

\begin{document}                

\title{Propagating beliefs in spin glass models}

\author{Yoshiyuki Kabashima}

\address{Department of Computational Intelligence and Systems Science, \\
Tokyo Institute of Technology, Yokohama 2268502, Japan \\
{kaba@dis.titech.ac.jp}
} %

\begin{abstract}                
We investigate dynamics of an inference algorithm termed 
the belief propagation (BP) when employed in spin glass (SG) models 
and show that its macroscopic behaviors can be traced by recursive 
updates of certain auxiliary field distributions whose stationary 
state reproduces the replica symmetric solution offered 
by the equilibrium analysis. We further provide a compact 
expression for the instability condition of the BP's fixed point
which turns out to be identical to that of 
instability for breaking the replica symmetry in equilibrium 
when the number of couplings per spin is infinite. 
This correspondence is extended to a SG model of finite connectivity 
to determine the phase diagram, which is numerically supported. 
\end{abstract}
\maketitle

Recently, there is growing interest in a similarity between 
researches on spin glass (SG) and information processing (IP) 
\cite{Nishimori}. 
Since employment of methods from SG theory
provided significant progresses for several problems related to IP 
such as machine learning \cite{learning}, 
error-correcting \cite{Sourlas,us,Nishimori_Wong,Montanari}
and spreading codes \cite{Tanaka,Kaba_CDMA}, 
it is natural to expect that the opposite direction might be possible. 

The purpose of this article is to show such an example. 
More specifically, we show that investigating dynamics of an iterative 
inference algorithm termed the belief propagation (BP) 
which has been developed in IP research \cite{Pearl,MacKay} 
when employed in SG models provides a new understanding about 
thermodynamical properties of SG. 
We show that the replica symmetric (RS) solution known 
in the equilibrium analysis can be characterized 
as a {\em macroscopically} stationary state in BP. 
We also provide a compact expression of the 
{\em microscopic} instability condition around the fixed point 
in the BP dynamics which turns out to be identical to 
that of instability for breaking the replica symmetry in equilibrium 
termed the Almeida-Thouless (AT) instability \cite{AT} when the 
number of connectivity per spin is infinite. 
Efficacy of this expression for a sparsely connected 
SG model is also numerically supported.

We here take up a family of Ising SG models define by Hamiltonian  
\begin{eqnarray}
{\cal H}(\bS|\bJ)=-\sum_{\mu=1}^M J_{\mu} \prod_{l \in {\cal L}(\mu)}S_l, 
\label{eq:hamiltonian}
\end{eqnarray}
where ${\cal L}(\mu)$ denotes a set of indices which are connected to a 
quenched coupling $J_\mu$. We assume that each coupling is independently 
generated from an identical distribution 
\begin{eqnarray}
P(J_\mu)=\frac{1+J_0/(\sqrt{C}J)}{2} 
\delta\left (J_\mu-\frac{J}{\sqrt{C}} \right )
+\frac{1-J_0/(\sqrt{C}J)}{2} \delta\left (J_\mu+\frac{J}{\sqrt{C}} \right ). 
\label{eq:coupling}
\end{eqnarray}
We further assume that for each $\mu$, ${\cal L}(\mu)$ is composed of randomly 
selected $K \sim O(1)$ spin indices and each spin index $l$ is concerned with 
$C$ couplings the set of which is denoted as ${\cal M}(l)$. 
$J_0>0$ and $J>0$ are parameters to control 
the mean and the standard 
deviation of $J_\mu$, respectively, which naturally links the current system 
(\ref{eq:hamiltonian}) to the Sherrington-Kirkpatrick 
(SK) model \cite{SK} in the case of $K=2$ and $C \sim O(N)$ and 
to sparsely connected SG models \cite{WS,KS1,Murayama_Okada} in general.

\begin{figure}[t]
\setlength{\unitlength}{1mm}
\begin{center}
\begin{picture}(136,30)
\epsfxsize=70mm
\put(-10,-5){
\epsfbox{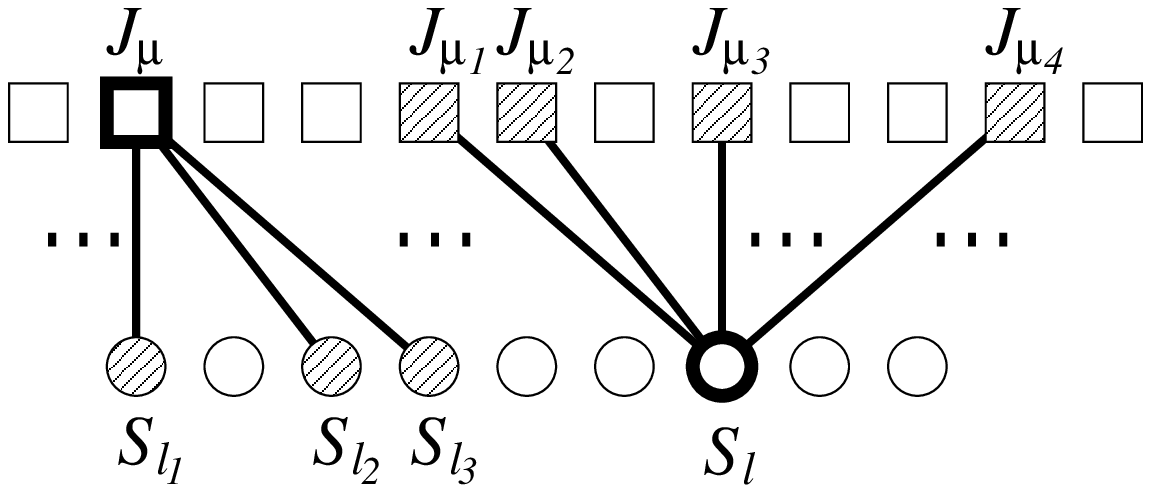}
}
\put(70,1.5){
\epsfbox{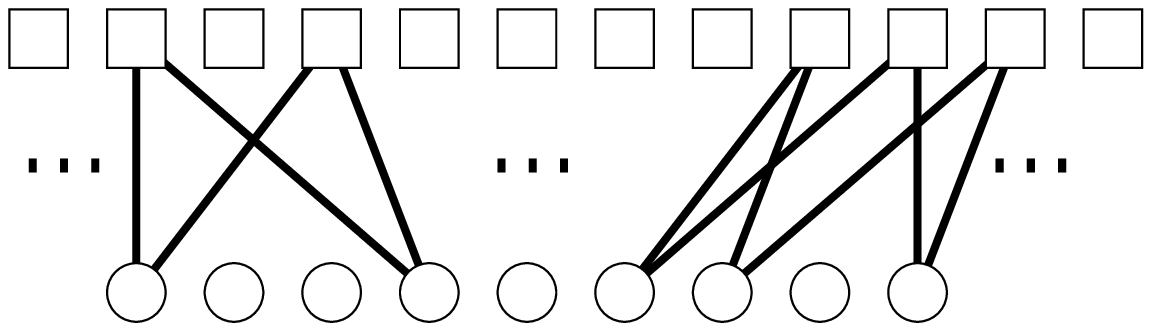}}
\put(0,30){(a)}
\put(80,30){(b)}
\end{picture}
\end{center}
\caption{
(a): Graphical expression of SG models in the case of $K=3$ and 
$C=4$.  In this expression, each spin $S_l$ denoted as $\bigcirc$ 
is linked to $C=4$ couplings $J_\mu$ ($\Box$), each 
of which is connected to $K=3$ spins. ${\cal L}(\mu)$ 
and ${\cal M}(l)$ represent sets of indices of spins and 
couplings that are related to $J_\mu$ and $S_l$, respectively. 
In the figure, ${\cal L}(\mu)=\left \{ l_1,l_2,l_3 \right \}$ and 
${\cal M}(l)=\left \{ \mu_1,\mu_2,\mu_3,\mu_4 \right \}$. 
(b): Cycles in a graph. A cycle is composed of multiple 
paths to link an identical pair of nodes. 
It is shown that BP can provide the exact spin averages
in a practical time scale if a given graph is free from 
cycles \cite{Pearl}. 
}
\label{fig:fig1}
\end{figure}

A major goal of statistical mechanics in the current system
is to calculate the microscopic spin average 
$m_l=\mathop{\rm Tr}_{\bS} S_l \exp \left [ -\beta {\cal H}(\bS|\bJ) \right ]/
\mathop{\rm Tr}_{\bS} \exp \left [ -\beta {\cal H}(\bS|\bJ) \right ]$
from given Hamiltonian (\ref{eq:hamiltonian}). 
This is formally identical to an inference problem for 
a posterior distribution 
$P(\bS|\bJ) \propto \prod_{\mu=1}^M P(J_\mu|\bS)$ derived 
from a conditional probability $P(J_\mu|\bS) = 
\exp\left [\beta J_\mu \prod_{l \in {\cal L}(\mu)}S_l \right ]/
\sum_{J_\mu=\pm J/\sqrt{C}} 
\exp\left [\beta J_\mu \prod_{l \in {\cal L}(\mu)}S_l \right ]$ 
and a uniform prior, which 
can be expressed in a bipartite graph as Figure \ref{fig:fig1} (a). 
In this expression, spins and couplings are
denoted as two different types of nodes and are linked by edges
when they are directly connected, which is useful to 
explicitly represent statistical dependences between 
estimation variables (spins) and observed data (couplings). 

BP is an iterative algorithm defined over the bipartite graph 
to calculate the spin average for a given set of couplings $\bJ=(J_\mu)$
\cite{Pearl,MacKay}. 
In the current system, this is performed by passing {\em beliefs} (or messages)
between the two types of nodes via edges at each update as
\begin{eqnarray}
\hat{m}_{\mu l}^{t+1} &=& \tanh \beta J_\mu 
\prod_{k \in {\cal L}(\mu) \backslash l} m_{\mu k}^t, \label{eq:horizontal} \\
m_{\mu l }^t &=&\tanh \left ( \sum_{\nu \in {\cal M}(l) \backslash \mu} 
\tanh^{-1} \hat{m}_{\nu l}^{t} \right ), \label{eq:vertical}
\end{eqnarray}
where beliefs $m_{\mu l }^t$ and $\hat{m}_{\mu l }^t$ are parameters to 
represent auxiliary distributions 
at $t$th  update as
$P(S_l|\{J_{\nu \ne \mu }\})=(1+m_{\mu l}^t S_l)/2$ 
and
$P(J_\mu|S_l, \{J_{\nu \ne \mu }\})=\mathop{\rm Tr}_{S_{k \ne l}}
P(J_\mu |\bS)P(\bS|\{J_{\nu \ne \mu }\}) \propto 
(1+\hat{m}_{\mu l}^t S_l)/2$, respectively. 
${\cal L}(\mu) \backslash l$ stands for a set of spin indices 
which belong to ${\cal L}(\mu)$ other than $l$ and 
similarly to ${\cal M}(l) \backslash \mu$. 
Calculating $\hat{m}_{\mu l}$ iteratively, 
the estimate of the spin average at $t$th update is provided as 
\begin{eqnarray}
m_l^t =\tanh \left ( \sum_{\mu \in {\cal M}(l)} \tanh^{-1} \hat{m}_{\mu l}^t 
\right ). 
\label{eq:t_th_average}
\end{eqnarray}

It is known that BP provides the exact spin average 
by the convergent solution when the bipartite graph 
is free from cycles (Figure \ref{fig:fig1} (b)). Actually, 
BP is a very similar scheme to the transfer matrix method (TMM) or the 
Bethe approximation \cite{Bethe,KS} which is frequently 
used in physics and the current statement can be regarded as 
a generalization of a known property of TMM that 
offers the exact results for a one dimensional lattice or a tree. 
However, BP still has a possibility to introduce something new 
into physics since it is explicitly expressed as an {\em algorithm} and 
such view point has been rare in the research on matters. 
This strongly motivates us to examine its dynamical properties, 
which we will focus on hereafter.

Let us first discuss the macroscopic behavior of the BP dynamics
(\ref{eq:horizontal}) and (\ref{eq:vertical}). Although the 
current randomly constructed system is not free from cycles, 
it can be shown that the typical length of the cycles 
grows as $O(\ln N)$ with respect to the system size $N$
as long as $C$ is $O(1)$ \cite{Renato},
which implies that the self-interaction from 
the past state is presumably negligible in the thermodynamic limit. 
On the other hand, the self-interaction is also expected as sufficiently 
small even if $C$ is large since the strength of the coupling becomes weak as
$O(C^{-1/2})$. This and eqs. (\ref{eq:horizontal}) and 
(\ref{eq:vertical}) imply that the time evolution of 
the macroscopic distributions of 
beliefs $\pi^t(x) \equiv(1/NC)\sum_{l=1}^N \sum_{\mu \in {\cal M}(l)} 
\delta(x-m_{\mu l}^t)$ and 
$\hat{\pi}^t(\hat{x}) \equiv (1/NC)\sum_{l=1}^N 
\sum_{\mu \in {\cal M}(l)} \delta(\hat{x}-\hat{m}_{\mu l}^t)$
is likely to be well captured by recursive equations
\begin{eqnarray}
\hat{\pi}^{t+1}(\hat{x})&=&\int \prod_{l=1}^{K-1} dx_l \pi^t(x_l)
\left \langle \delta
\left (\hat{x}-\tanh \beta {\cal J} \prod_{l=1}^{K-1}x_l \right )
\right \rangle_{{\cal J}}, \label{eq:pi}\\
\pi^{t}({x})&=&\int \prod_{\mu=1}^{C-1} d\hat{x}_\mu \hat{\pi}^t(\hat{x}_\mu)
\delta\left (x-\tanh \left (\sum_{\mu=1}^{C-1}\tanh^{-1} \hat{x}_\mu \right )
\right ), \label{eq:pi_hat}
\end{eqnarray}
where $\left \langle \cdots \right \rangle_{{\cal J}}$
represents the average with respect to ${\cal J}$ following 
distribution (\ref{eq:coupling}). 

The validity of the current argument and its link to the replica 
symmetric (RS) ansatz in the equilibrium analysis 
have been shown already for finite $C$ \cite{Richardson,Renato2}. 
Here, we further show that these can be extended to the case of 
infinite $C$ even if the AT stability of the RS solution is broken 
in equilibrium. 

When $C$ becomes infinite, it is more convenient to deal with 
an auxiliary field of finite strength 
$h^t_{\mu l} \equiv \sum_{\nu \in {\cal M}(l)\backslash \mu} 
\tanh^{-1} \hat{m}^t_{\nu l}\simeq 
\sum_{\nu \in {\cal M}(l)\backslash \mu} 
\hat{m}^t_{\nu l}$ rather than $\hat{m}^t_{\nu l}$ since
$\hat{m}^t_{\mu l}$ becomes infinitesimal. 
Due to the central limit theorem, the distribution of the 
auxiliary field $\rho^t(h)\equiv (1/NC) 
\sum_{\l=1}^N \sum_{\mu \in {\cal M}(l)}
\delta(h-h^t_{\mu l})$ 
can be regarded as a Gaussian 
\begin{eqnarray}
\rho^t(h)=\int \prod_{\mu=1}^{C-1} 
d \hat{x}_{\mu} \hat{\pi}^t(\hat{x}_{\mu} ) 
\delta(h-\sum_{\mu=1}^{C-1} \tanh^{-1} \hat{x}_\mu)
\simeq 
\frac{1}{\sqrt{2 \pi F^t}}\exp \left [ -\frac{(h-E^t)^2}{2F^t} \right ], 
\label{eq:gaussian}
\end{eqnarray}
where $E^t$ and $F^t$ are the average and the variance to parameterize
the Gaussian distribution $\rho^t(h)$, respectively. 
The expression in the middle 
implies $\pi^t(x)=\int dh \rho^t(h)\delta(x-\tanh(h))$. Plugging this 
into eq. (\ref{eq:pi}) and recursively employing eq. (\ref{eq:gaussian}), 
we obtain a compact expression for the update of $E^t$ and $F^t$ as
\begin{eqnarray}
E^{t+1}&=&\beta J_0 \left (M^t \right)^{K-1}, \quad F^{t+1}=\beta^2 J^2 
\left( Q^t \right )^{K-1}, \label{eq:EF_MQ} \\
M^{t}&=&\int D z \tanh (\sqrt{F^t}z +E^t ), \quad Q^{t}=\int D z 
\tanh^2 (\sqrt{F^t}z +E^t ), \label{eq:MQ_EF}
\end{eqnarray}
where $Dz \equiv \exp[-z^2/2]/\sqrt{2 \pi} $ and
$M^t$ and $Q^t$ can be expressed as 
$M^t \simeq (1/N)\sum_{l=1}^N m_{\mu l}^t \simeq (1/N)\sum_{l=1}^N m_{l}^t $
and $Q^t \simeq (1/N)\sum_{l=1}^N (m_{\mu l}^t)^2 
\simeq (1/N)\sum_{l=1}^N (m_{l}^t)^2$, respectively, 
due to the law of large numbers. 
Eqs. (\ref{eq:EF_MQ}) 
and (\ref{eq:MQ_EF}) serve as alternatives 
of eqs. (\ref{eq:pi}) and (\ref{eq:pi_hat}). 

\begin{figure}[t]
\setlength{\unitlength}{1mm}
\begin{center}
\begin{picture}(136,50)
\epsfxsize=100mm
\put(-20,-5){
\epsfbox{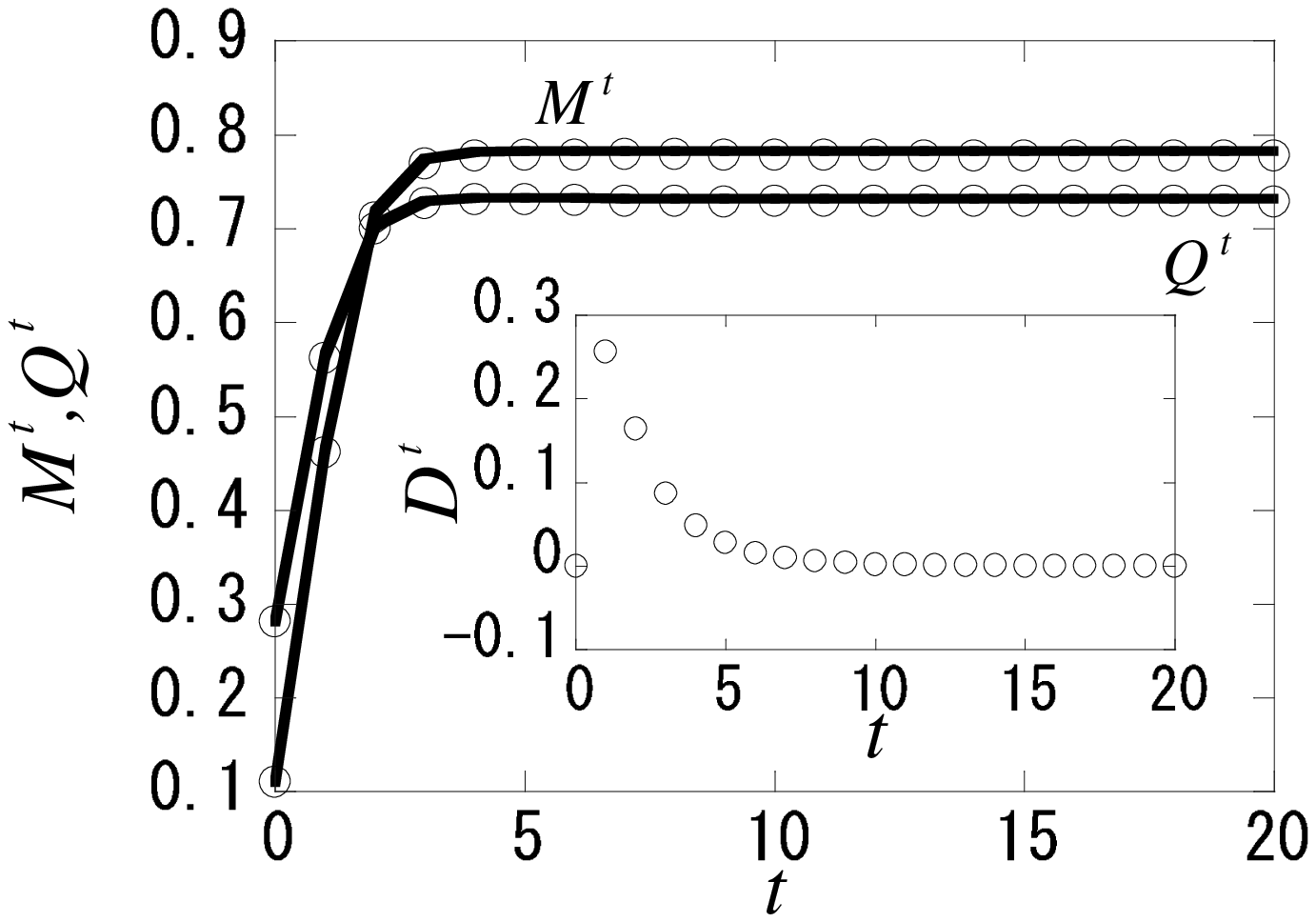}
}
\put(60,-5){
\epsfbox{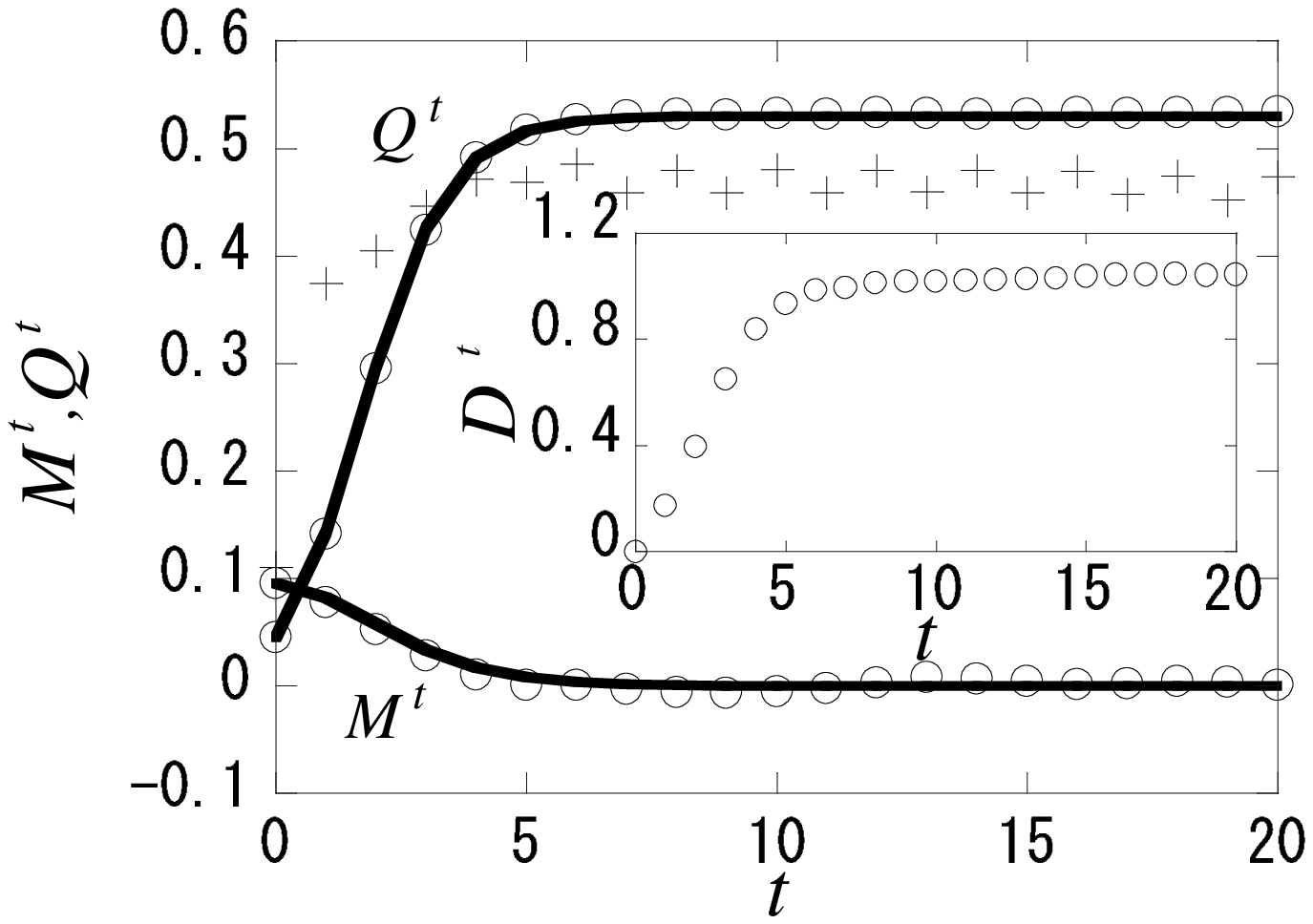}}
\put(0,50){(a)}
\put(80,50){(b)}
\end{picture}
\end{center}
\caption{
Time evolution of macroscopic variables
\protect$M^t=(1/N)\sum_{l=1}^N m_l^t$ and 
\protect$Q^t=(1/N) \sum_{l=1}^N (m_l^t)^2$ in 
the SK model for the BP updates (\ref{eq:horizontal}) 
and (\ref{eq:vertical}) ({\bf BP}: $\bigcirc$),  
the reduced dynamics (\ref{eq:EF_MQ}) and (\ref{eq:MQ_EF}) ({\bf RS}: lines) 
and the naive iteration of the TAP equation (\ref{eq:gen_TAP}) 
({\bf TAP}: \protect$+$) for (a) $J_0=1.5$ and (b) $J_0=0.5$ keeping 
$J=1$ and $T=0.5$. {\bf TAP} is plotted only for 
$Q^t$ in the case of $J_0=0.5$ 
in order to save space. Each marker is obtained from 
100 experiments for $N=1000$ systems. 
The AT stability is satisfied for $J_0=1.5$ but broken for $J_0=0.5$. 
Irrespectively of the AT stability, the behavior of the macroscopic 
variables in the BP dynamics can be well captured by the 
reduced dynamics while the naive iteration of the TAP equation does not 
exhibit any convergence even in the macroscopic scale. 
Insets: Squared deviation of spin averages between the successive updates 
$D^t=(1/N)\sum_{l=1}^N (m_l^t-m_l^{t-1})^2$ 
is plotted for the BP dynamics. The deviation vanishes to zero 
indicating convergence to a fixed point solution for $J_0=1.5$ 
while remains finite signalling instability of the fixed point 
for $J_0=0.5$. 
}
\label{fig:fig2}
\end{figure}

It should be noticed here that these equations can be regarded as 
the forward iteration of the saddle point equations to obtain 
the RS solution in the replica analysis of the multi-spin 
interaction infinite connectivity SG models \cite{Nishimori} 
and, in particular, of the SK model for $K=2$ \cite{SK}. 
In order to confirm the validity of the above argument, 
we compared the time evolution of the belief 
update (\ref{eq:horizontal}) and (\ref{eq:vertical}) ({\bf BP}) 
with that of eqs. (\ref{eq:EF_MQ}) and (\ref{eq:MQ_EF}) ({\bf RS})
for the SK ($K=2$) model, which is shown in Figures \ref{fig:fig2} (a)
and (b). 
We also compared them with the trajectory of 
the naive iteration of the BP's fixed point condition 
\begin{eqnarray}
m_l=\tanh \left (\sum_{\mu \in {\cal M}(l)} \! \beta 
J_\mu \! \prod_{k \in {\cal L}(\mu)
\backslash l} m_k \! - \! \sum_{\mu \in {\cal M}(l)} \! (\beta J_\mu)^2 \! 
\sum_{j \in {\cal L}(\mu) \backslash l} \! 
\left ( \! \prod_{k \in {\cal L}(\mu)
\backslash l,j}  \! m_k \right )^2(1-m_j^2) m_l \right ), 
\label{eq:gen_TAP}
\end{eqnarray}
({\bf TAP}) which can be obtained inserting $m_{\mu l} \simeq m_l -(1-m_l^2) 
\hat{m}_{\mu l}$ to the fixed point of 
eqs. (\ref{eq:horizontal}) and (\ref{eq:vertical}) $m_{\mu l}^{t}=m_{\mu l}$, 
$\hat{m}_{\mu l}^t =\hat{m}_{\mu l}$ and $m_{l}^t=m_l$.  
This becomes identical to the famous Thouless-Anderson-Palmer (TAP) 
equation of the SK model, in particular, for $K=2$ \cite{TAP}. 

The experiments were performed for $J_0=1.5$, $0.5$ keeping 
$J=1$ and $T=0.5$, where the AT stability of the RS solution 
in equilibrium is satisfied for $J_0=1.5$ but broken for 
$J_0=0.5$ \cite{AT}. 
Figures \ref{fig:fig2} (a) and (b) show that 
{\bf BP} and {\bf RS} exhibit excellent consistency with 
respect to the macroscopic variables irrespectively 
of whether the AT stability is satisfied or not. 
This strongly validates the reduction from BP 
(\ref{eq:horizontal}) and (\ref{eq:vertical}) to the macroscopic dynamics
(\ref{eq:EF_MQ}) and (\ref{eq:MQ_EF}). 
On the other hand, {\bf TAP} is considerably 
different from the others. 
In a sense, this may be natural because naively iterating 
eq. (\ref{eq:gen_TAP}) is just one of procedures 
for obtaining a solution and its trajectory in dynamics does not necessarily 
have any consistency with {\bf BP} or {\bf RS} while 
the BP's fixed point is correctly characterized by the TAP 
equation (\ref{eq:gen_TAP}) which does have a certain 
relation to the RS solution in equilibrium as shown in \cite{MPV}. 
These figures also imply that the dynamics of BP cannot be traced by a closed
set of equations with respect to singly indexed variables 
$m_l^t$ even for $C \to \infty$ while the fixed point condition 
in this limit is provided as coupled equations of $m_l$ (\ref{eq:gen_TAP}), 
which is also observed in a similar system \cite{Kaba_CDMA}.

Although Figures \ref{fig:fig2} (a) and (b) show that the 
macroscopic variables rapidly converge to those of 
the RS solution in BP, this does not imply that BP microscopically 
converges to a certain solution. 
In order to probe this microscopic convergence, we 
examined the squared difference of spin averages between 
successive updates $D^t \equiv (1/N) \sum_{l=1}^N (m_l^{t}-m_l^{t-1})^2$, 
the time evolution of which is shown in the insets of Figures 
\ref{fig:fig2} (a) and (b). 
These illustrate that the (microscopic) 
local stability of the BP's fixed point 
can be broken even if the macroscopic behavior seems to converge,  
which cannot be detected by only examining the reduced macroscopic 
dynamics (\ref{eq:EF_MQ}) and (\ref{eq:MQ_EF}). 

In order to characterize such instability, we next turn to 
the stability analysis of the BP updates (\ref{eq:horizontal}) and 
(\ref{eq:vertical}). Linearizing the updates with respect to 
the auxiliary field $h_{\mu l}=\tanh^{-1} m_{\mu l}$ 
around a fixed point solution $m_{\mu l}^t=m_{\mu l}$, we obtain a 
dynamics of the auxiliary field fluctuation $\delta h_{\mu l}^t$ as
\begin{eqnarray}
\delta h_{\mu l}^{t+1} 
=\sum_{\nu \in {\cal M}(l) \backslash \mu}
\frac{\tanh \beta J_\nu \prod_{k \in {\cal L}(\nu)\backslash l} 
m_{\nu k}}
{1- \left (\tanh \beta J_\nu \prod_{k \in {\cal L}(\nu)\backslash l} 
m_{\nu k} \right )^2 }
\times \sum_{j \in {\cal L}(\nu ) \backslash l} 
\frac{1-m_{\nu j}^2}{m_{\nu j}} \times \delta h_{\nu j}^t. 
\label{eq:fluctuation}
\end{eqnarray}
Analytically solving this linearized equation 
for a large graph is generally difficult. 
However, since the current system is randomly constructed, 
the self-interaction of $\delta h_{\mu l}^t$ from the past 
can be considered as small as those of beliefs are. 
This implies that the time evolution of the fluctuation distribution 
$f^t(y)\equiv (1/NC) \sum_{l=1}^N \sum_{\mu \in {\cal M}(l)}
\delta(y-\delta h_{\mu l}^t)$ can be provided by a functional equation 
\begin{eqnarray}
f^{t+1}(y)&=&\prod_{\mu=1}^{C-1}\prod_{l=1}^{K-1}
d y_{\mu l}f^t(y_{\mu l})  \cr
&\times &
\left \langle
\delta 
\left (
y-\sum_{\mu=1}^{C-1} \frac{\tanh \beta 
{\cal J}_\mu \prod_{k=1}^{K-1} x_{\mu k}}
{1-\left (\tanh \beta {\cal J}_\mu \prod_{k=1}^{K-1} x_{\mu k} \right )^2 }
\times \sum_{l=1}^{K-1} \frac{1-x_{\mu l}^2}{x_{\mu l}} \times y_{\mu l}
\right )
\right \rangle_{{\cal J}_\mu, x_{\mu l}}, 
\label{eq:fluctuation_prop}
\end{eqnarray}
where $\left \langle \cdots \right \rangle_{{\cal J}_\mu, x_{\mu l}}$
denotes the average over ${\cal J}_\mu$ and $x_{\mu l}$ following 
eq. (\ref{eq:coupling}) and the stationary distribution 
of $\pi^t(x)=\pi(x)$, 
respectively, and the stability of the BP's fixed 
point can be characterized by whether the stationary solution 
$f^t(y)=f(y)=\delta(y)$ 
is stable or not in update (\ref{eq:fluctuation_prop}).
This formulation makes analytical investigation 
possible to a certain extent. 

In order to connect eq. (\ref{eq:fluctuation_prop}) 
to the existing analysis, let us first investigate the limit $C \to  \infty$
for which much more results are known compared to the case of finite $C$. 
Due to the central limit theorem, the distribution of the 
field fluctuation can be assumed as a Gaussian 
$f^t(y)=(1/\sqrt{2 \pi b^t}) \exp\left [ -(y-a^t)^2/(2 b^t) \right ]$, 
where $a^t$ and $b^t$ are the mean and the variance of 
the distribution, respectively. 
Plugging this expression into eq. 
(\ref{eq:fluctuation_prop}) offers update rules with
respect to $a^t$ and $b^t$ as
$a^{t+1}=\beta J_0 M^{K-2}(1- Q) a^t$ and 
$b^{t+1}=(\beta J)^2 Q^{K-2} 
\int Dz \left (1-\tanh^2 (\sqrt{F} z+E) \right )^2\left (b^t
+(a^t)^2 \right )$, 
where $M,Q,E$ and $F$ represent the convergent solutions of eqs. 
(\ref{eq:EF_MQ}) and (\ref{eq:MQ_EF}). 
In order to examine the stability of $f(y) = \delta(y)$, 
we linearize these equations around $a^t=b^t=0$, which 
provides the critical condition of the instability 
with respect to the growth of $b^t$
\begin{eqnarray}
(\beta J)^2 Q^{K-2} \int Dz \left (1-\tanh^2 (\sqrt{F} z+E) \right )^2=1, 
\label{eq:AT}
\end{eqnarray}
which becomes identical to that of the AT stability 
for the infinite range multi-spin interaction 
SG models and, in particular, for the SK model when $K=2$ \cite{AT}. 
Furthermore, in the case of the SK model ($K=2$), 
the critical condition with respect to 
$a^t$ around the paramagnetic solution $M=Q=0$ corresponds 
to the para-ferromagnetic transition. These mean that the two different 
phase transitions from the paramagnetic solution can be linked in a 
unified framework to the dynamical instabilities of BP 
by eq. (\ref{eq:fluctuation_prop}). 

When $C$ is finite, one can numerically perform the stability analysis 
employing eq. (\ref{eq:fluctuation_prop}), 
the detail of which will be reported elsewhere.
In addition, analytical investigation becomes possible 
for $K=2$ as follows since transitions from 
the paramagnetic solution in this case occur 
due to the {local} instability. 

For a small $\beta$, the paramagnetic solution 
$\pi(x)=\hat{\pi}(x)=\delta(x)$ ($m_{\mu l}=\hat{m}_{\mu l}=0$)
expresses the correct stable fixed point of the BP dynamics. 
Inserting this into eq. (\ref{eq:fluctuation_prop}) does not 
provide a closed set of equations with respect to a finite number 
of parameters since $f^t(y)$ is no more a Gaussian. 
However, assuming $f^t(y)\simeq \delta(y)$, 
the stability analysis can be reduced to coupled equations
with respect to the mean and the variance of $f^t(y)$ as 
$a^{t+1}=(C-1)\left \langle \tanh \beta {\cal J} \right \rangle_{\cal J}
a^t$ and $b^{t+1}=(C-1) \left (
\left \langle \tanh^2 \beta {\cal J} \right \rangle_{\cal J} 
b^t + \left (\left \langle \tanh^2 \beta {\cal J} \right \rangle_{\cal J} 
- \left \langle \tanh \beta {\cal J} \right \rangle_{\cal J}^2 \right )
(a^t)^2 \right )$. Linearizing these around $a^t=b^t=0$ 
provides the critical conditions with respect to the 
growth of $a^t$ and $b^t$ as 
\begin{eqnarray}
(C-1)\left \langle \tanh \beta {\cal J} \right \rangle_{\cal J}&=&1, 
\label{eq:ferro_sparse} \\
(C-1)\left \langle \tanh^2 \beta {\cal J} \right \rangle_{\cal J}&=&1, 
\label{eq:SG_sparse}
\end{eqnarray}
respectively. It should be mentioned that a similar condition to 
eq. (\ref{eq:SG_sparse}) was once obtained for a SG model on 
the Bethe lattice \cite{Thouless} 
while eq. (\ref{eq:ferro_sparse}) was not. 
However, the current scheme may be superior to that employed 
in \cite{Thouless} as the expression (\ref{eq:fluctuation_prop}) 
is compact and, therefore, 
can be easily extended to the case of multi-spin interaction ($K \ge 3$)
with the aid of numerical methods while such extension 
requires higher order perturbation and becomes highly complicated 
in the other scheme. 

\begin{figure}[t]
\setlength{\unitlength}{1mm}
\begin{center}
\begin{picture}(150,70)
\epsfxsize = 150mm
\put(0,-5){
\epsfbox{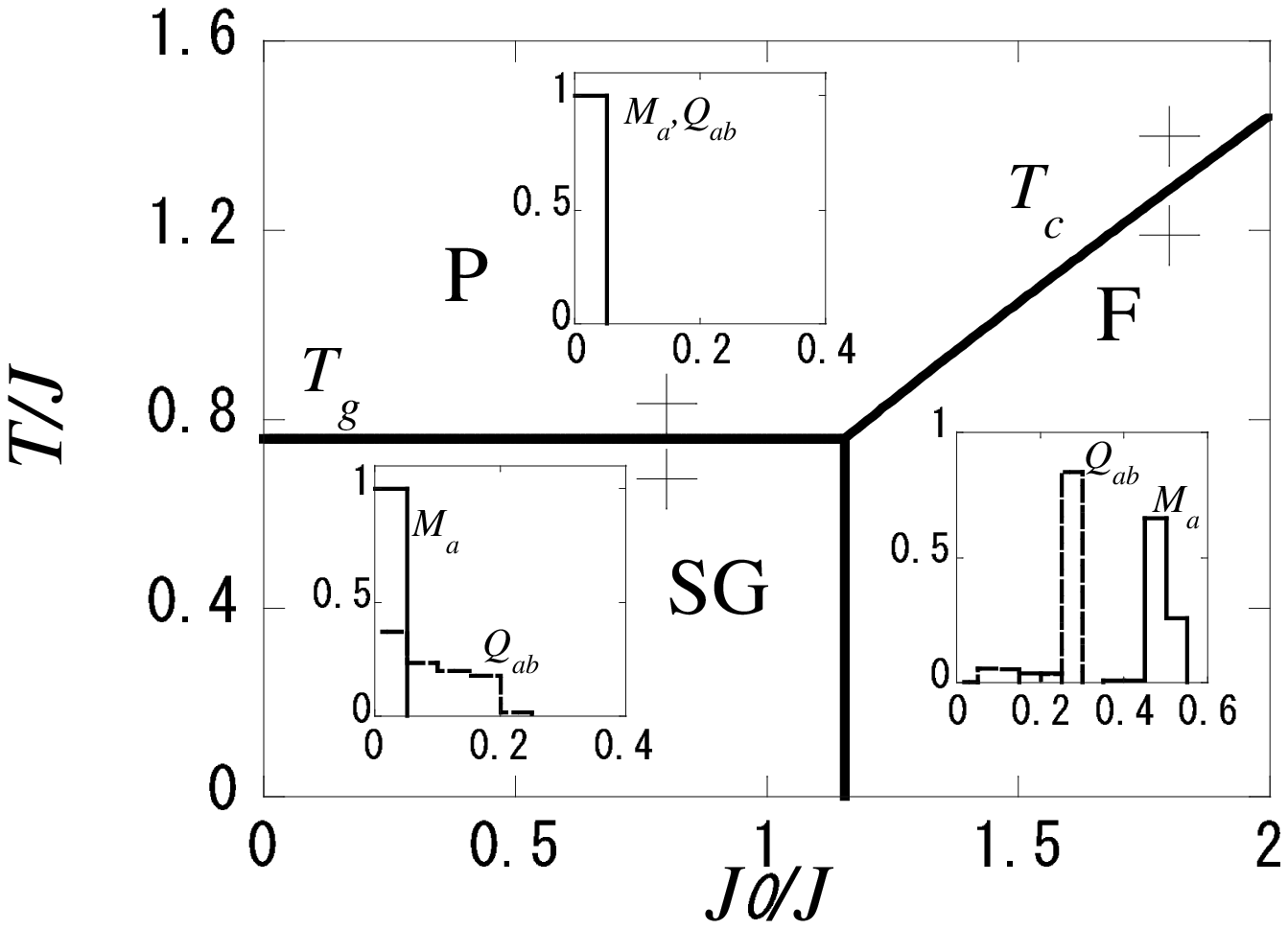}
}
\end{picture}
\end{center}
\caption{
Phase diagram for $K=2$ and $C=4$ suggested by 
eqs. (\ref{eq:ferro_sparse}) and (\ref{eq:SG_sparse}). 
P, F and SG stand for the paramagnetic, 
the ferromagnetic and the spin glass phases, respectively. 
The boundary between F and SG is just a conjecture. 
In order to examine the validity of this diagram, 
100 Monte Carlo experiments were performed for $N=2000$ systems 
at conditions denoted by $+$. 
For each condition, frequencies of macroscopic 
magnetizations $M_a=(1/N) \sum_{l=1}^N m_l^a$ and 
overlaps $Q_{ab}=(1/N)\sum_{l=1}^N m_l^a m_l^b$ $(a>b)$
were evaluated, where $m_l^a$ is the average of $S_l$ 
obtained from 20000 Monte Carlo steps per spin
for experiments $a,b=1,2,\ldots,100$ (insets). 
For both of the two conditions in P, 
all of $M_a$ and $Q_{ab}$ fall into the first bin. 
On the other hand, sharp peaks indicate the 
order to the ferromagnetic state in F and a broad 
distribution of $Q_{ab}$ signals 
the breaking of the replica symmetry in SG. }
\label{fig:fig3}
\end{figure}

Eqs. (\ref{eq:ferro_sparse}) and (\ref{eq:SG_sparse}) 
might correspond to the para-ferromagnetic and 
the para-SG phase transitions, respectively, since they do in the limit 
$C \to \infty$. In order to examine this, 
we performed numerical experiments for $N=2000$ and $C=4$. Although 
further investigation may be necessary to prove correctness, 
the data obtained from $100$ experiments of $20000$ Monte Carlo steps per
spin exhibit good consistency with 
the analytical expressions (\ref{eq:ferro_sparse}) 
and (\ref{eq:SG_sparse}) 
indicating that the correspondence between the phase transitions 
in equilibrium and the dynamical instabilities of BP holds 
for finite $C$ as well (Figure \ref{fig:fig3}). 

In summary, we have investigated dynamical behavior of BP when employed
in SG models. We have shown that the time evolution of macroscopic 
variables can be well captured by recursive updates of auxiliary 
field distributions which becomes identical to 
the forward iteration of the saddle point equations under the RS ansatz
in the replica analysis. We have further shown that the dynamical 
instability of the BP's fixed point is closely related to the AT 
instability of the RS solution, which has been numerically supported.

Relationship between the current scheme and an 
existing AT analysis for finite connectivity 
SG models \cite{Mottishaw} that generally 
requires complicated calculation and is not 
frequently employed in practice is under investigation. 
Besides this, extension of the current framework to the 
replica symmetry breaking (RSB) schemes \cite{Parisi,Mezard_Parisi} 
is a challenging and interesting future work. 

This work was partially supported by Grants-in-Aid from 
the MEXT, Japan, Nos. 13680400, 13780208 and 14084206.

\vfil\eject

\end{document}